\documentclass[sigconf,screen]{acmart}

\usepackage{wrapfig}
\AtBeginDocument{%
  }

\setcopyright{acmlicensed}
\copyrightyear{2018}
\acmYear{2018}
\acmDOI{XXXXXXX.XXXXXXX}

\acmConference[FSE Companion '25]{Make sure to enter the correct
  conference title from your rights confirmation emai}{June 23--27,
  2025}{Trondheim, Norway}
\acmBooktitle{Companion Proceedings of the 33rd ACM Symposium on the Foundations of Software Engineering (FSE '25), June 23--27, 2025, Trondheim, Norway}
\acmISBN{978-1-4503-XXXX-X/18/06}

\begin{document}

\title{Alibaba LingmaAgent: Improving Automated Issue Resolution via Comprehensive Repository Exploration}

\author{Yingwei Ma, Qingping Yang, Rongyu Cao, Binhua Li, Fei Huang, Yongbin Li}
\email{{mayingwei.myw, yangqingping.yqp, caorongyu.cry, binhua.lbh, f.huang, shuide.lyb}@alibaba-inc.com}
\affiliation{%
  \institution{Tongyi Lab, Alibaba Group}
  \country{}
  }

\renewcommand{\shortauthors}{Ma et al.}

\begin{abstract}
This paper presents Alibaba LingmaAgent, a novel Automated Software Engineering method designed to comprehensively understand and utilize whole software repositories for issue resolution. Deployed in TONGYI Lingma, an IDE-based coding assistant developed by Alibaba Cloud, LingmaAgent addresses the limitations of existing LLM-based agents that primarily focus on local code information. Our approach introduces a top-down method to condense critical repository information into a knowledge graph, reducing complexity, and employs a Monte Carlo tree search based strategy enabling agents to explore and understand entire repositories. We guide agents to summarize, analyze, and plan using repository-level knowledge, allowing them to dynamically acquire information and generate patches for real-world GitHub issues. In extensive experiments, LingmaAgent demonstrated significant improvements, achieving an 18.5\% relative improvement on the SWE-bench Lite benchmark compared to SWE-agent. In production deployment and evaluation at Alibaba Cloud, LingmaAgent automatically resolved 16.9\% of in-house issues faced by development engineers, and solved 43.3\% of problems after manual intervention. Additionally, we have open-sourced a Python prototype of LingmaAgent for reference by other industrial developers~\footnote{\url{https://github.com/RepoUnderstander/RepoUnderstander}}. \textit{In fact, LingmaAgent has been used as a developed reference by many subsequently agents.}

\end{abstract}

\ccsdesc[500]{Software and its engineering~Automatic programming}
\ccsdesc[500]{Computing methodologies~Multi-agent planning}

\keywords{Automatic Software Engineering (ASE), Software Engineering Agents, Large Language Models (LLMs), Fault Localization, Automated Program Repair, Monte Carlo Tree Search (MCTS)}


\maketitle

\section{Introduction}
\label{sec-intro}
Automated Software Engineering (ASE) explores the automation of complex software development processes and develops innovative tools to improve software lifecycle.
Recent years, in the ASE domain, LLM-based agents have demonstrated their strong general abilities, e.g., the environment awareness ability~\citep{hong2023metagpt, wang2024codeact, kong2024contrastrepair}, planning \& reasoning ability~\citep{cognitionai2023devin, opendevin2023, luo2024repoagent, wang2024codeact}, tool construction~\citep{zhang2024codeagent} ability, etc. 

More recently, an exemplary milestone termed Devin~\citep{cognitionai2023devin} explores an end-to-end LLM-based agent system for complex real-world SE tasks (i.e., fix real-world Github issues). It plans user requirements, utilizes editor and terminal tools for independent decision-making and reasoning, and eventually generates code patches to meet the needs. This innovative approach has garnered considerable attention from the AI and SE communities~\citep{ACR, yang2024sweagent}. For instance, SWE-agent~\citep{yang2024sweagent} strategically designs an Agent Computer Interface (ACI) to empower SE agents in creating \& editing code files, navigating repositories, and executing programs. Additionally, AutoCodeRover~\citep{ACR} extracts abstract syntax trees in programs, iteratively searches for useful information based on requirements, and generates program patches.

Although these works achieved promising performance, their designs, focusing on local code information, failed to grasp the global context and intricate interdependencies among functions and classes. 
For example, SWE-agent~\citep{yang2024sweagent} maintains a context window within a code file that allows the agent to scroll up and down. AutoCodeRover~\citep{ACR} searches functions or classes within the whole repository. 
Typically, the code comprising a full logic chain for a specific functionality is not arranged sequentially within a single file; rather, it is logically scattered across multiple folders and files.
It is difficult to retrieve all relevant code files among maybe thousands of files in a repository, especially starting only from the text in user requirements.
This paper argues that a comprehensive understanding of the whole repository becomes the most critical path to ASE. This also is the basis for the multi-file editing functions in existing commercial software like Cursor~\citep{cursor} and TONGYI Lingma~\citep{Lingma}.

Undoubtedly, it is challenging to utilize the vast information of an entire repository within LLM. 
Firstly, a GitHub repository may contain thousands of code files, making it impractical to include them all in the context windows of LLM. Even if it could, an LLM would struggle to accurately capture the code relevant to the objective within such an extensive context.
Secondly, the intrinsic logic of how the code execution is distinctly different from the sequence of the code text in a file. For instance, the location where a bug triggers an error message and the actual place that requires modification may not be in the same file, yet they are certainly logically connected.

To address these challenges, we propose LingmaAgent, a novel ASE method deployed in TONGYI Lingma (short for Lingma). Lingma is an IDE-based coding assistant recently developed by Alibaba Cloud and available to users worldwide. 
Inspired by how human software engineers approach project-level issues, LingmaAgent guides LLM-based agents to first gain a comprehensive understanding of the entire software repository. This approach enables agents to grasp the overall structure and dependencies, thereby enhancing their ability to effectively resolve issues within the broader context of the project.

Specifically, we construct a repository knowledge graph using a top-down approach, organizing the repository into a hierarchical structure tree that provides a clear understanding of code context and scope. This structure is further enhanced by expanding it into a reference graph, capturing intricate function call relationships and facilitating comprehensive dependency and interaction analysis.
Subsequently, we propose a Monte Carlo Tree Search (MCTS) based repository exploration method. Specifically, the agents first collect the critical information regarding to the SE task on the repository knowledge graph by the explore-and-exploit strategy. Then, by simulating multiple trajectories and evaluating their reward score, our method iteratively narrows down the search space and guide the agents to focus on the most relevant areas. 
In addition, to better utilize the repository-level knowledge, we guide the agents to summarize, analyze, and plan for the repository information.  Finally, the agents are instructed to manipulate the search API tools to dynamically acquire local information, and fix the real-world issues by generating patches.

We demonstrate the superiority and effectiveness of LingmaAgent through extensive experiments and comprehensive analyses. Using the SWE-bench benchmark \citep{jimenez2024swebench}, we evaluate our method's capabilities for issue resolution. Our experiments reveal an 18.5\% relative improvement compared to SWE-agent on the SWE-bench Lite benchmark. In production deployment and evaluation at Alibaba Cloud, LingmaAgent automatically resolved 16.9\% of in-house issues faced by development engineers and solved 43.3\% of problems after manual intervention.

The main contributions of this paper are summarized as follows.

\begin{itemize}
\item We highlight the whole repository understanding as the crucial path to ASE and propose a novel agent-based method named LingmaAgent to solve the challenges. 

\item We propose to condense the extensive codes and complex relations of the repository into the knowledge graph in a top-to-down mode, improving performance and efficiency. 

\item We design a Monte Carlo tree search based repository exploration strategy to assist the comprehensive understanding of the whole repository for the issue-solving agents. 

\item Extensive experiments and analyses demonstrate the superiority and effectiveness of LingmaAgent.
\end{itemize}

\section{Related Work}


\subsection{LLM-based Software Engineering Agents}

In recent years, Large Language Model (LLM) based AI agents have advanced the development of automatic software engineering. AI agents improve the capabilities of project-level software engineering (SE) tasks through running environment awareness~\citep{hong2023metagpt, wang2024codeact, kong2024contrastrepair}, planning \& reasoning~\citep{wang2024codeact, cognitionai2023devin, opendevin2023, luo2024repoagent}, and tool construction~\citep{zhang2024codeagent, lee2024unified}. Surprisingly, Devin~\citep{cognitionai2023devin} is a milestone that explores an end-to-end LLM-based agent system to handle complex SE tasks. Concretely, it first plans the requirements of users, then adopts the editor, terminal and search engine tools to make independent decisions and reasoning, and finally generates codes to satisfy the needs of users in an end-to-end manner. Its promising designs and performance swiftly ignited unprecedented attention from the AI community and SE community to Automatic Software Engineering (ASE)~\citep{yang2024sweagent,ACR}. For example, SWE-agent~\citep{yang2024sweagent} carefully designs an Agent Computer Interface (ACI) to empower the SE agents capabilities of creating \& editing code files, navigating repositories, and executing programs. Besides, AutoCodeRover~\citep{ACR} extracts the abstract syntax trees in programs, then iteratively searches the useful information according to requirements, and eventually generates program patches. Their designs mainly focus on the local information in the repository, e.g., code files, classes, or functions themselves. Although achieving promising performance, from the insights of the human SE developers, the excellent understanding of the whole repository is a critical path to ASE.

\subsection{Evaluation of LLM-based Software Engineering Agents}
Benefiting from the strong general capability of LLMs, LLM-based software engineering agents can handle many important SE tasks, e.g., repository navigation~\citep{zhang2024codeagent, wang2024codeact}, code generation~\citep{hong2023metagpt, ding2024cycleagent1, ishibashi2024selfagent2, tang2024collaborativeagent3, rasheed2024codeporiagent4}, debugging~\citep{hong2023metagpt, yang2024sweagent, ACR}, program repair~\citep{qin2024agentfl, ACR, yang2024sweagent}. The existing methods usually regard code generation as a core ability and mainly conduct evaluations on it. Precisely, the code generation test set~\citep{humaneval, mbpp, liu2024your, zheng2023codegeex, lu2021codexglue} consists of the short problem description, the solution, and the corresponding unit test data. However, with the fast development of LLMs and agents, these datasets are no longer able to comprehensively evaluate their capabilities in the real-world SE tasks. To this end, the repository-level code completion and generation tasks~\citep{liu2023repobench,ding2024crosscodeeval, du2024evaluating} are presented to evaluate the repository understanding and generation capabilities of LLMs and agents. More recently, SWE team\citep{jimenez2024swebench, yang2024sweagent} develop a unified dataset named SWE-bench to evaluate the capability of the agent system to solve real-world GitHub issues automatically. Specifically, it collects the task instances from real-world GitHub issues from twelve repositories. Consistent with previous evaluation methods, SWE-bench is based on the automatic execution of the unit tests. Differently, the presented test set is challenging and requires the agents to have multiple capabilities, including repository navigation, fault locating, debugging, code generation and program repairing. Besides, SWE-bench Lite~\citep{SWElite} is a subset of SWE-bench, and it has a similar diversity and distribution of repositories as the original version. Due to the smaller test cost and more detailed filtering, SWE-bench Lite is officially recommended as the benchmark of LLM-based SE agents. Therefore, consistent with previous methods~\citep{yang2024sweagent,ACR,opendevin2023}, we report our performance on SWE-bench Lite.

\subsection{Repository-level Code Intelligence}

With the development of AI technology, the field of code intelligence has gradually transitioned from solving single function-level or code snippet-level problems to real-world software development at the repository level. In the repository-level code intelligence task, there are many works~\citep{liang2024repofuse, ding2022cocomic, zhang2023repocoder, lozhkov2024starcoder, guo2024deepseek, zan2023private, shrivastava2023repository, bairi2023codeplan} that aim to leverage the large amount of code available in current repositories to help code models generate better, more accurate code. 
Among them, StarCoder2~\citep{lozhkov2024starcoder} and Deepseek-Coder~\citep{guo2024deepseek} model repository knowledge in the pre-training stage, sort repository files according to reference dependencies, and guide the model to learn the global dependencies of repository information. RepoCoder~\citep{zhang2023repocoder} continuously retrieves relevant content by iterating RAG, while methods such as CoCoMIC~\citep{ding2022cocomic} and RepoFuse~\citep{liang2024repofuse} jointly use the RAG module and the current file's dependency relationship module to introduce it into the context of LLM. Although the above methods enhance the model's understanding of the repository context to a certain extent, the repository-level code often contains complex contextual call relationships, and the RAG method alone may not be able to recall all semantically relevant content. In addition, there may be a large amount of complex irrelevant information in the RAG results, which interferes with the model's accurate fault location. Therefore, starting from the practical experience of software engineering, we simulated people's global experience in understanding the repository and experience-guided exploration and location to achieve more effective repository understanding.

\section{Methodology}
\subsection{Overview}

\begin{figure*}
    \centering
    \includegraphics[scale=0.113]{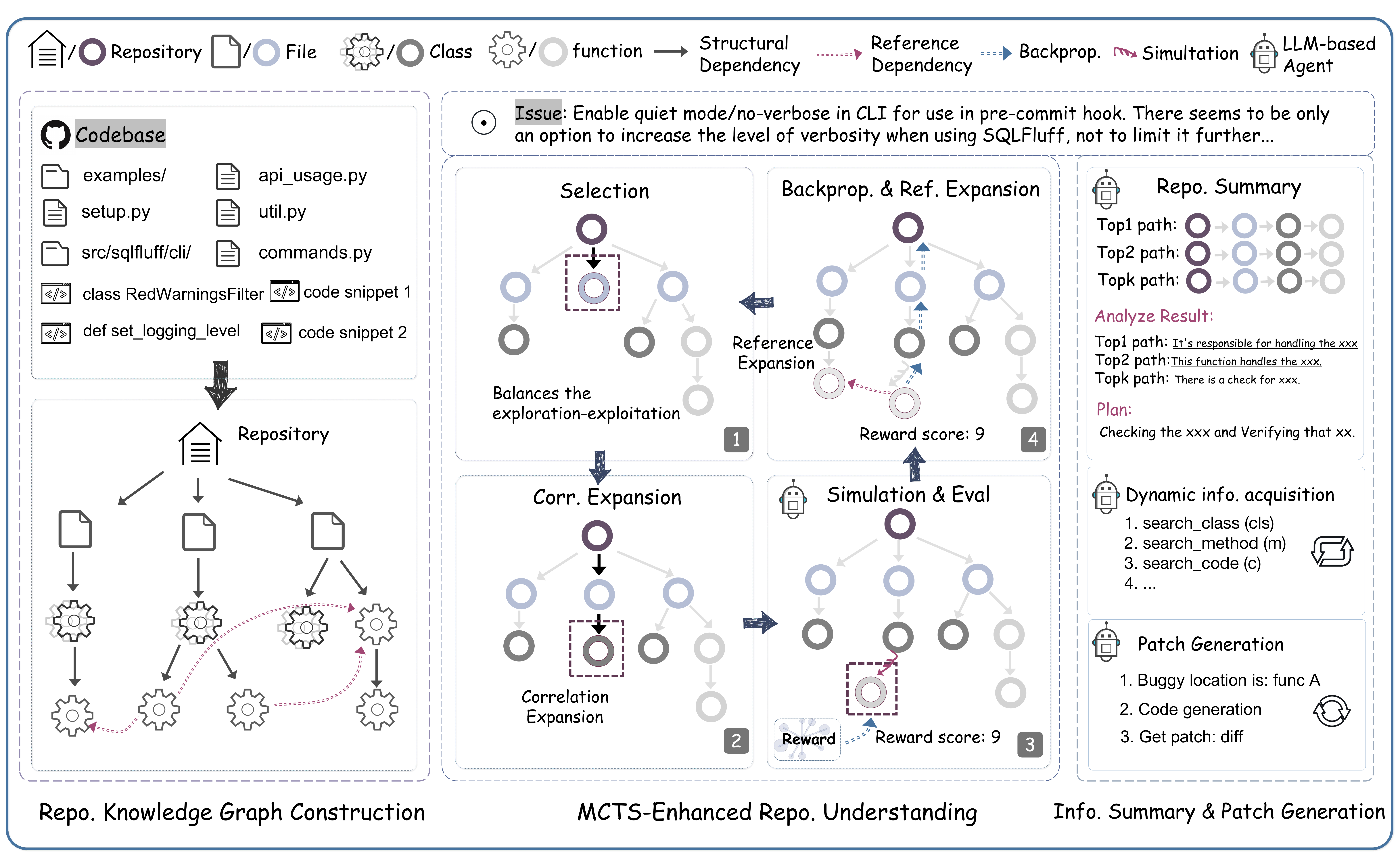}
    \caption{The overview of our proposed LingmaAgent. Firstly, we construct the repository knowledge graph is constructed to efficiently represent the code and the dependency in the repository. Subsequently, we empower the agents with the ability of repository understanding by designing the Monte Carlo tree search based repository explore strategy. In addition, we guide the agents to summarize, analyze, and plan to better utilize the repository-level knowledge. Then, they can manipulate the tools to dynamically acquire issue-relevant code information and generate the patches to solve the real-world GitHub issues.}
    \label{fig:overview}
\end{figure*}

We first describe the overall operating process of LingmaAgent, and introduce the stages in detail in the subsequent parts of this section. Given a workspace, LingmaAgent can automatically solve real-world issues. Among them, LingmaAgent involves three key steps, repository knowledge graph construction stage, MCTS-enhanced repository understanding stage, information utilization \& patch generation stage. The overall workflow is shown in Figure \ref{fig:overview}.

In \textbf{Repository Knowledge Graph Construction} phase, LingmaAgent first builds a repository knowledge graph to represent the entire repository and describe the relationships between entities. This is achieved by parsing the software structure and analyzing it in a top-down manner.  The repository is first organized into a hierarchical tree that allows a clear understanding of the context and scope of the code.  To facilitate comprehensive dependency and interaction analysis, the tree structure is further extended into a reference graph that captures function call relationships.

Due to the large scale and information complexity of the repository knowledge graph, during the \textbf{MCTS-Enhanced Repository Understanding} phase, LingmaAgent uses the Monte Carlo tree search algorithm to dynamically explore the entire graph. This method focus on discovering key information (i.e., repository functionality and dependency structure) that has a significant impact on issue solving. Through correlation expansion and reference expansion, MCTS simulates multiple trajectories and evaluates their importance, dynamically narrowing the search space and allocating computing resources to the most relevant regions.  This targeted navigation enables the model to efficiently access and process important information in the repository, thereby facilitating precise fault localization and informed patch generation.

Inspired by the actual development experience of human programmers, it is necessary to have certain global prior knowledge of the repository before solving specific tasks. Therefore, in \textbf{Information Utilization \& Patch Generation} phase, LingmaAgent first summarizes the important information collected in the repository understanding phase to form an experience of the entire repository. Then, in order for LingmaAgent to use the global experience to obtain dynamic information during the problem solving process, we follow AutoCodeRover\citep{ACR}'s information retrieval method and use API search tools to extract information in the repository knowledge graph. This includes specific classes and functions and code snippets, etc., to maintain local dynamic knowledge during the task. After collecting enough context, LingmaAgent uses global experience to summarize the currently acquired information to locate faults, generate modified code and return patches that try to resolve the issue.

\subsection{Repository Knowledge Graph Construction}

For human programmers, when solving project-level issues, developers first need to carefully review and understand the project's software repository to ensure that they have a full understanding of the functional modules and dependencies that may be involved. This includes building the hierarchical tree structure and call graph of the software repository. Through the hierarchical tree structure, developers can clearly see the overall architecture of the project and the relationship between each module; through the call graph, developers can understand the calling relationships and dependency paths between functions to identify the root causes of problems and  the potential impact of changes. 

Therefore, in order to learn from the practices of human programmers in understanding and maintaining code, we represent the entire repository as a repository knowledge graph and describe the relationships between entities by parsing the software structure~ (see Repo. Knowledge Graph Construction in Figure \ref{fig:overview}). First, we top-down analyze the structure of the software repository, organizing the repository into a hierarchical structure tree (including files, classes, and functions) to clearly understand the context and scope of the code. We then extend the tree structure into a reference graph containing function call relationships, allowing the model to perform comprehensive dependency and interaction analysis. Different from existing methods\citep{luo2024repoagent, ding2022cocomic}, our reference relationship only involves functions, because functions are the basic unit of program execution, and the calling relationship between functions directly affects the behavior and execution logic of the program. Excessive reference relationships may increase the complexity of the graph structure and affect the analysis efficiency and accuracy of the model. This structured repository knowledge graph not only improves the efficiency of the model in retrieving relevant information, but also ensures the consistency and reliability of the automated process.

Specifically, we recursively traverse each code file in the repository, use abstract syntax trees to parse the corresponding files respectively, and obtain basic units such as classes and functions, including their names, code snippets, paths, and locations in the files. We then add these elements to the structure tree from top to bottom. Finally, we analyze the calling relationship between functions and add corresponding directed edges to the graph. This in-depth understanding provides LLM agents with the necessary background knowledge and contextual information, allowing them to more accurately locate the problem and come up with effective solutions.

\subsection{MCTS-Enhanced Repository Understanding}

After building a repository knowledge graph, a comprehensive understanding of the information in the graph is critical to effectively solving problems. However, given the complexity and size of modern software systems, often containing hundreds of files and thousands of functions. The vast magnitude of the search space in large software repositories makes exhaustive analysis impractical. Furthermore, context length limitations of language models limit the amount of information that can be efficiently processed at given conversation. Therefore, without targeted methods to identify highly relevant nodes and edges in graphs, models may struggle to perform accurate and efficient analysis, hampering their ability to solve real-world software engineering problems.

To address these challenges, we propose an repository exploration approach that leverages Monte Carlo Tree Search (MCTS) to enhance LLM and agents' understanding of software repositories (see MCTS-Enhanced Repo. Understanding in Figure \ref{fig:overview}). This method systematically explores the repository knowledge graph and prioritizes the discovery of critical information such as repository functions and dependency structures that have a greater impact on resolving issues. By simulating multiple trajectories and evaluating their importance, MCTS dynamically narrows the search space and focuses computational resources on the most relevant areas. This targeted navigation enables models to access and process important information more efficiently, thus facilitating precise fault localization and informed patch generation. The MCTS process begins from a root node, representing the repository node, and unfolds in four iterative stages: selection, correlation expansion, simulation\&evaluation and backpropagation\&reference expansion. Below we describe each stage in further detail.

{\bfseries Selection.} The selection phase aims to balance exploration and exploitation problems in the node selection process. The main challenge in this phase is to maintain a balance between in-depth analysis of highly relevant content in the repository and a broad search for potentially important information throughout the repository. Delving excessively into high-correlation modules can cause the model within a local optimal solution, ignoring that other critical paths or dependencies may exist. Extensive search may lead to the dispersion of computing resources and the processing of a large amount of irrelevant information, which increases the burden on the model and reduces search efficiency. To balance the needs of the above two aspects, we use the UCT algorithm~\citep{kocsis2006bandit} for node selection, following the formula: $ UCT = \frac{w_i}{n_i} + c \sqrt{\frac{2 \ln n_p}{n_i}} $, where $w_i$ is the total reward of child node $i$. The calculation of specific rewards will be introduced in detail in Simulation \& Evaluation section. $n_i$ is the number of visits to child node $i$ and $n_p$ is the number of visits to the parent node. $c$ is the exploration parameter used to adjust the balance between exploration and exploitation. In this work, we set $c$ to $\sqrt{2}/2$.

{\bfseries Correlation Expansion.} During the expansion process, leaf nodes are expanded to incorporate new nodes. If the current leaf node has a child node in the repository knowledge graph, the most likely child node is selected instead of random expansion. In this stage, we designed two methods: Correlation expansion and Reference relationship expansion. In this section, we mainly introduce correlation expansion, and reference relationship expansion will be introduced in the Backpropagation \& Reference Expansion section. Similar code is most likely to be code related to user requirements. User requirements or issues usually contain some keywords that may add new or modified functions. Therefore, we use the bm25 score to calculate the relevance~\citep{ding2024crosscodeeval, husain2019codesearchnet, xie2023survey}, and give priority to codes with higher relevance for expansion. Correlation expansion can effectively match user requirements with relevant nodes in the software knowledge graph, thereby improving the accuracy and efficiency of node expansion.

{\bfseries Simulation \& Evaluation.} After completing the expansion, we enter the simulation process. During the simulation, we start from the newly expanded node and simulate along possible paths to evaluate the effectiveness of these paths in solving the current issue. Consistent with the correlation expansion method, we continuously and recursively select the child nodes with the highest correlation scores in the software knowledge graph until leaf nodes, and then reward the nodes. 

In the evaluation phase, we need to evaluate the relevance of the selected leaf nodes to the issue, including classes, top-level functions, class methods or sub-functions, etc. However, traditional evaluation methods usually rely on keyword matching and semantic matching algorithms, which perform poorly when dealing with complex software systems and diverse problem descriptions.

\begin{figure}
    \centering
    \includegraphics[scale=0.27]{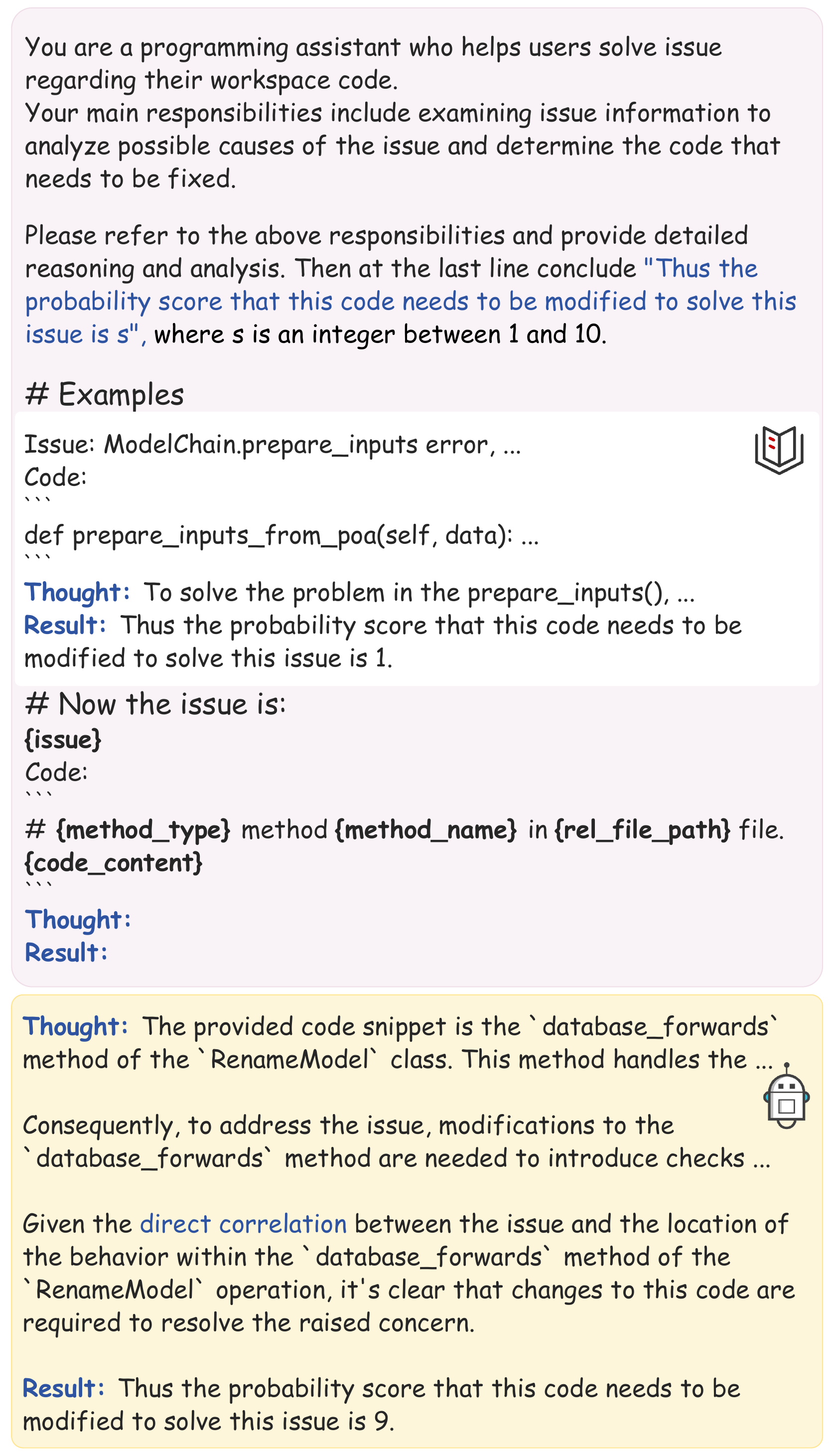}
    \caption{Reward agent's input prompt template and output results, with some details omitted.}
    \label{fig:prompt_reward}
\end{figure}

Drawing upon previous work on in-context learning (ICL) and Chain-of-Thought (CoT)~\citep{dong2022survey, workrethinking, wei2022cot}, we employ a reward method based on ICL and CoT to provide reward scores. Our approach leverages the advanced ability of LLMs to learn and optimize reward functions from limited examples of programming practice to accurately assess the correlation between leaf nodes and problem descriptions. Specifically, we first use ICL to let the language model learn to understand the core functions and operating modes of the reward function in a given context. Then, the CoT is used to enable the model to conduct in-depth reasoning based on the specific information in the question and code snippets to evaluate the correlation of leaf nodes. The reward function prompt template we designed (see Figure \ref{fig:prompt_reward}) starts with a guided system prompt that clearly points out the goals and responsibilities of the reward function. Then, through a series of example combinations of \textit{<issue description, code snippets, thinking process, results>}, the input, output and reasoning chain in the scoring process are demonstrated. Finally, the prompt ends with a new set of issue descriptions and code snippets, at which point the model is expected to learn the intermediate reasoning steps from the given examples and output corresponding reward scores. Finally, we only keep the nodes with a reward score of no less than 6 and return their content and structural dependencies.

Compared with traditional methods, our method reduces the dependence on large amounts of labeled data. This is critical to cope with diverse and evolving situations in software repository, as traditional approaches may suffer from the limitations of labeled data. Therefore, our method has better adaptability and accuracy when resolving real-world software development environments.

{\bfseries Backpropagation \& Reference Expansion.} 
After the evaluation ends, we perform a bottom-up update from the terminal node back to the root node. During this process, we update the visit count $n$ and the reward value $w$. In addition, we also introduced reference relationship expansion in the backpropagation phase. Different from the conventional expansion method, we not only expand when we encounter leaf nodes, but also when we encounter those nodes with higher reward scores (set the threshold to a reward score of not less than 6 here), we will expand their reference modules and objects based on the repository knowledge graph. And then integrate them into new nodes. The insight is that in actual development, the node called by the current node is often the key node for function implementation, and the called node is usually the use of the current node and depends on the implementation and changes of the current node. Therefore, if a node has a higher reward score, the nodes with calling relationships may also be relevant. By expanding these calling relationship nodes, code snippets related to the current issue can be captured more comprehensively.

\subsection{Information Utilization \& Patch Generation}

At this stage, LingmaAgent first summarizes the whole repository experience, then obtains code snippet information dynamicly on this basis, and finally generates patches that try to solve the problem. The three steps are detailed below.

\begin{figure}
    \centering
    \includegraphics[scale=0.27]{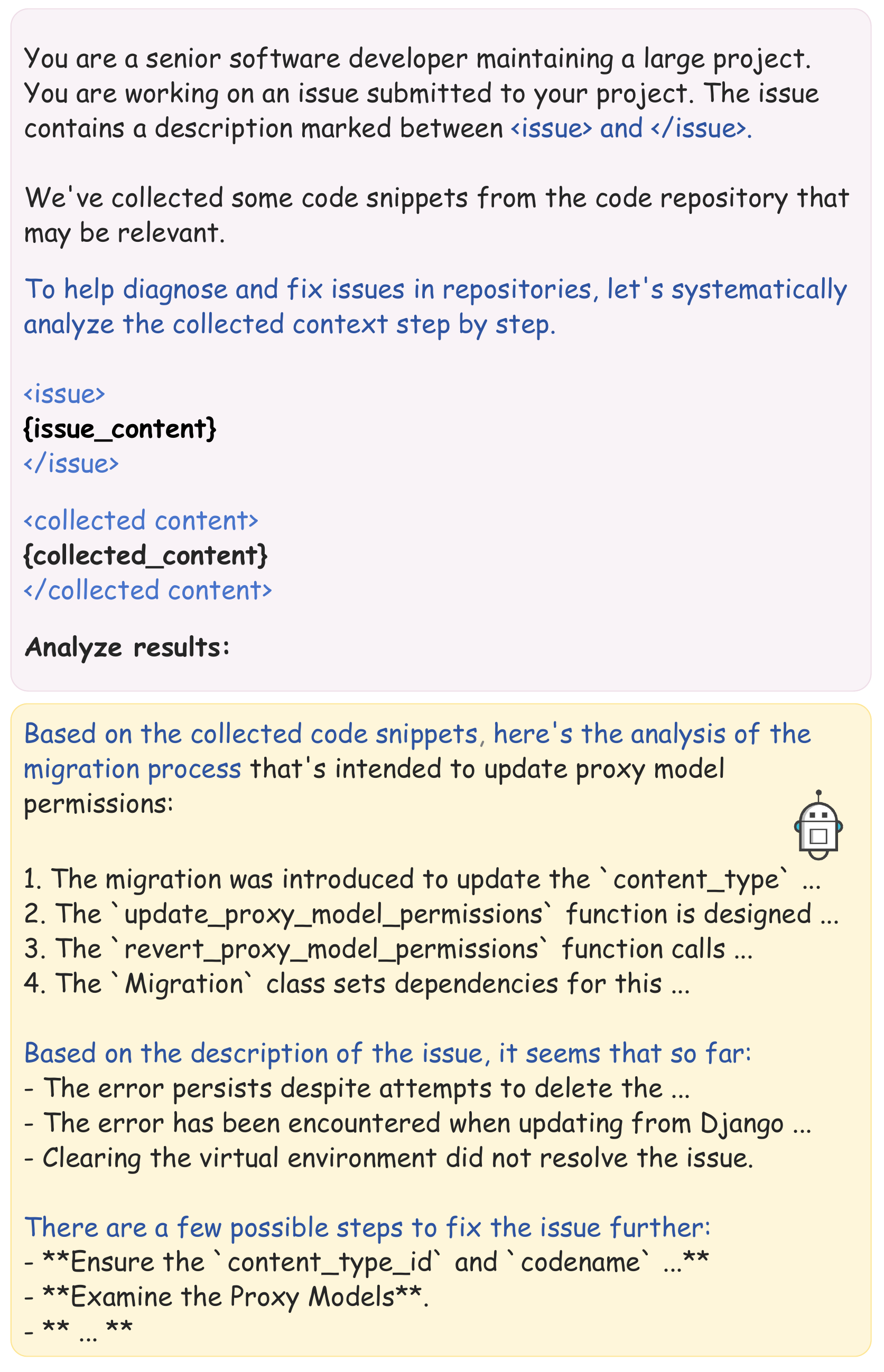}
    \caption{Summary agent's input prompt template and output results, with some details omitted.}
    \label{fig:prompt_summary}
\end{figure}

{\bfseries Repository Summary.} To more effectively utilize the global repository information collected during the repository understanding phase, we introduce a summary agent. The agent aims to systematically analyze and summarize the code snippets collected in the repository knowledge graph and submitted issues , and then plan how to solve the problem, thereby forming an experience of the entire repository. Specifically, the summary agent takes the issue and the collected relevant code fragments as input, and then outputs a summary of the relevant fragments in sequence and plans a solution. The specific prompt template is shown in Figure \ref{fig:prompt_summary}. Since the collected global repository information may be complex and contain a large number of code fragments and annotation descriptions, we only use the location description of the relevant code fragments (i.e., structural dependencies in the repository) and the output of the Summary Agent (i.e., summary and planning) as LingmaAgent’s experience to guide subsequent actions. This experience does not include specific function implementation, but only focuses on overall repository experience guidance. The location description is formalized as \textit{<file>a.py</file><class>Class A</class><func>func a</func>}, and the summary agent output is as shown in Figure \ref{fig:prompt_summary}.

{\bfseries Dynamic information acquisition.} Global experience information is LingmaAgent's experience summary of relevant information in the current workspace, which can help the language model understand issues and find solutions more quickly. In the process of solving problems, in order to make full use of this global experience information, LingmaAgent futher needs to dynamically extract local information from the current repository, including specific classes, functions and code snippets in the repository.

The ReAct~\citep{yao2022react} framework (i.e., Reson then Act) guides the model to generate inference trajectories and task-specific actions in a staggered manner, allowing the model to interact with the code repository and collect information. Specifically, the ReAct framework first generates reasoning paths through the Chain-of-Thought~\citep{wei2022cot}, and then outputs actual actions based on the reasoning results. Therefore, by using ReAct method, LingmaAgent can call the corresponding search API according to task requirements  and dynamically extract local information from the current repository to collect relevant context. We follow AutoCodeRover's search API method~\citep{ACR}, using the three-layer search method of \textit{search\_class}, \textit{search\_method}, and \textit{search\_code}. Specifically, LingmaAgent first independently determines the API that needs to be called. Then the retrieval API will search for classes, methods and code snippets in the repository knowledge graph, and finally return the results to the agent.

{\bfseries Patch Generation.} In the patch generation step, LingmaAgent first locates faults based on the summary of global experience and dynamic information, extracts the context of code snippets that may need to be modified, and then generates modified code snippets. Finally, a diff is generated based on the code snippet before modification and the code snippet after modification, and is returned as the final result. If a diff is incorrect due to syntax, we will retry until an applicable patch with correct syntax is generated. We follow AutoCodeRover~\citep{ACR} and set the maximum number of retries to 3 to ensure that the generated patch can be applied as much as possible.

\section{Experiment}

To validate the performance of LingmaAgent, we conduct a series of comprehensive experiments and comparisons. We begin by comparing LingmaAgent with RAG-based and Agent-based systems on the SWE-bench Lite dataset (§\ref{superiority}). We then assess the consistency and reliability of LingmaAgent's performance across multiple runs (§\ref{random}). In addition, we conduct detailed ablation studies to understand the contribution of each component in LingmaAgent (§\ref{ablation}). Finally, we assess LingmaAgent's effectiveness in industrial settings using an in-house dataset from Alibaba Cloud, testing both fully automated and human-in-the-loop scenarios (§\ref{inhouse}).

\subsection{Experimental Setup}\label{setup}

{\bfseries Datasets.} We evaluate on the SWE-bench Lite dataset~\citep{jimenez2024swebench} which are constructed due to the high cost of evaluating in the complete SWE-bench. SWE-bench Lite includes 300 task instances sampled from SWE-bench, following a similar repository distribution. SWE-bench team recommend future systems evaluating on SWE-bench to report numbers on SWE-bench Lite in lieu of the full SWE-bench set if necessary. SWE-bench Lite aims to provide a diverse set of code base issues that can be verified using in-repository unit tests. It requires LLM systems to generate corresponding patches based on the actual issues in the repository, and then pass the tests.

{\bfseries Baselines.} 
We compare LingmaAgent with two types of baselines. The first category is the RAG baselines~\citep{jimenez2024swebench}. This type of baseline uses the BM25 method to retrieve code base files related to the issue and inputs them into LLM to directly generate patch files that solve the problem. The second type of baseline is the agents baseline~(i.e., AutoCodeRover~\citep{ACR} and SWE-agent~\citep{yang2024sweagent}), which locates the problem through complex multiple rounds of interaction and execution feedback, and finally generates a patch to solve the problem through iterative verification.


{\bfseries Metrics.} 
Following the SWE-bench~\citep{jimenez2024swebench}, We evaluate the effectiveness of LingmaAgent, using the percentage of resolved instances and the patch application rate. Among them, the patch application rate refers to the proportion of instances where code changes are successfully generated and can be applied to existing code bases using Git tools. Resolved ratio represents the overall effectiveness of solving actual GitHub issues, and application ratio reflects the intermediate results of patch availability.

{\bfseries Configurations.} 
All results, ablations, and result analyzes of LingmaAgent use the GPT4-Turbo model (i.e., gpt-4-1106-preview~\citep{gpt4}, the same model with SWE-agent~\citep{yang2024sweagent}). We use ast\footnote{https://docs.python.org/3/library/ast.html} and Jedi\footnote{https://github.com/davidhalter/jedi} library to parse repository and obtain syntax structures and dependencies of repository. In MCTS-Enhanced Repository Understanding stage, we set the number of search iterations to 600 and maximum search time to 300 seconds. In information Utilization \& Patch Generation stage, we set the maximun number of summary code snippets to 10. SWE-bench has a relatively complex environment configuration. Thanks to the development of the open source community, we use the well-build open source docker of the AutoCodeRover~\citep{ACR} team for experiments.

\subsection{Comparison Experiment}\label{superiority}

We first evaluate the effectiveness of LingmaAgent in SWE-bench Lite (300 instances). The performance comparison analysis between LingmaAgent and other methods is shown in Table~\ref{tab:main_results}. In each instance, we provide a natural language description from a real-world software engineering problem and a local code repository of corresponding versions, asking the model to solve the problem and generate patches that can pass local automated testing. Resolved reflects the end-to-end ability of the current RAG LLM system and Agent system to solve software engineering problems. The results show that LingmaAgent is significantly better than other RAG and Agent systems, achieving SOTA performance on the test set. Compared with the RAG system, our method improves performance by nearly 5 times. Compared with the state-of-the-art Agent system, \textbf{we improve the accuracy of SWE-agent by 18.5\%}. These excellent performances demonstrate the advancement of our approach. In addition, the Apply application rate indicates the availability of generated patches. We found that Agent-based systems all achieved high availability, while RAG-based systems have lower availability, which proves that agent systems may be an important means to automatically solve software engineering tasks.

\begin{table}
\renewcommand{\arraystretch}{1.1}
\centering
\scalebox{1.0}{
\begin{tabular}{llcc}
\hline
\textbf{Method} & \textbf{Resolved} & \textbf{Apply} & \textbf{Avg Cost}\\
\hline
\textit{RAG-based} & & &\\
SWE-Llama 7B & 1.33\%  & 38.00\% & - \\
SWE-Llama 13B & 1.00\%  & 38.00\% & - \\
GPT-4	& 2.67\%  &	29.67\% & \$0.13 \\
Claude-3 Opus	& 4.33\%  &	51.67\% & \$0.25 \\\hline
\textit{Agent-based} & & & \\
AutoCodeRover & 16.11\%  & 83.00\% & \$0.45 \\
SWE-agent & 18.00\%  & 93.00\% & \$2.51 \\ 
\textbf{LingmaAgent} & \textbf{21.33\% (18.5\%{\color{red} $\uparrow$})} & 85.67\% &\$3.99\\ \hline
\end{tabular}}
\caption{Main results for LingmaAgent performance on the SWE-bench-lite test set. The numbers in brackets indicate the number of issues solved.}
\label{tab:main_results}
\end{table}

\begin{table}
\renewcommand{\arraystretch}{1.1}
\centering
\scalebox{0.9}{
\begin{tabular}{lccc}
\hline
\textbf{Method} & \textbf{Resolved} & \textbf{Apply} &\textbf{Avg Cost}\\
\hline
ACR \& SWE-agent & 24.33\% (73) & 98.00\% & - \\
Lingma \& ACR & 25.33\% (76) & 94.67\% & - \\ 
\textbf{Lingma \& SWE-agent} & \textbf{26.67\% (80)} & 99.67\% & - \\ \hline
\textbf{\textit{LingmaAgent (w.feedback)}} & & & \\

GPT-4 & 27.70\% (83) & 96.30\% & \$4.72 \\ 
GPT-4o & 28.33\% (85) & 95.33\% & \$2.39 \\
Claude3.5 Sonnet & 32.00\% (96) & 96.67\% & \$2.27 \\
\textbf{Claude3.5 Sonnet v1022} & \textbf{38.33\% (115) {\color{red} $\uparrow$}} & 98.00\% & \$2.18 {\color{blue} $\downarrow$ } \\
\hline
\end{tabular}}
\caption{Complementarity analysis of our method and baselines.}
\label{tab:venn_diagram_analysis}
\end{table}

\begin{figure}
    \centering
    \includegraphics[scale=0.25]{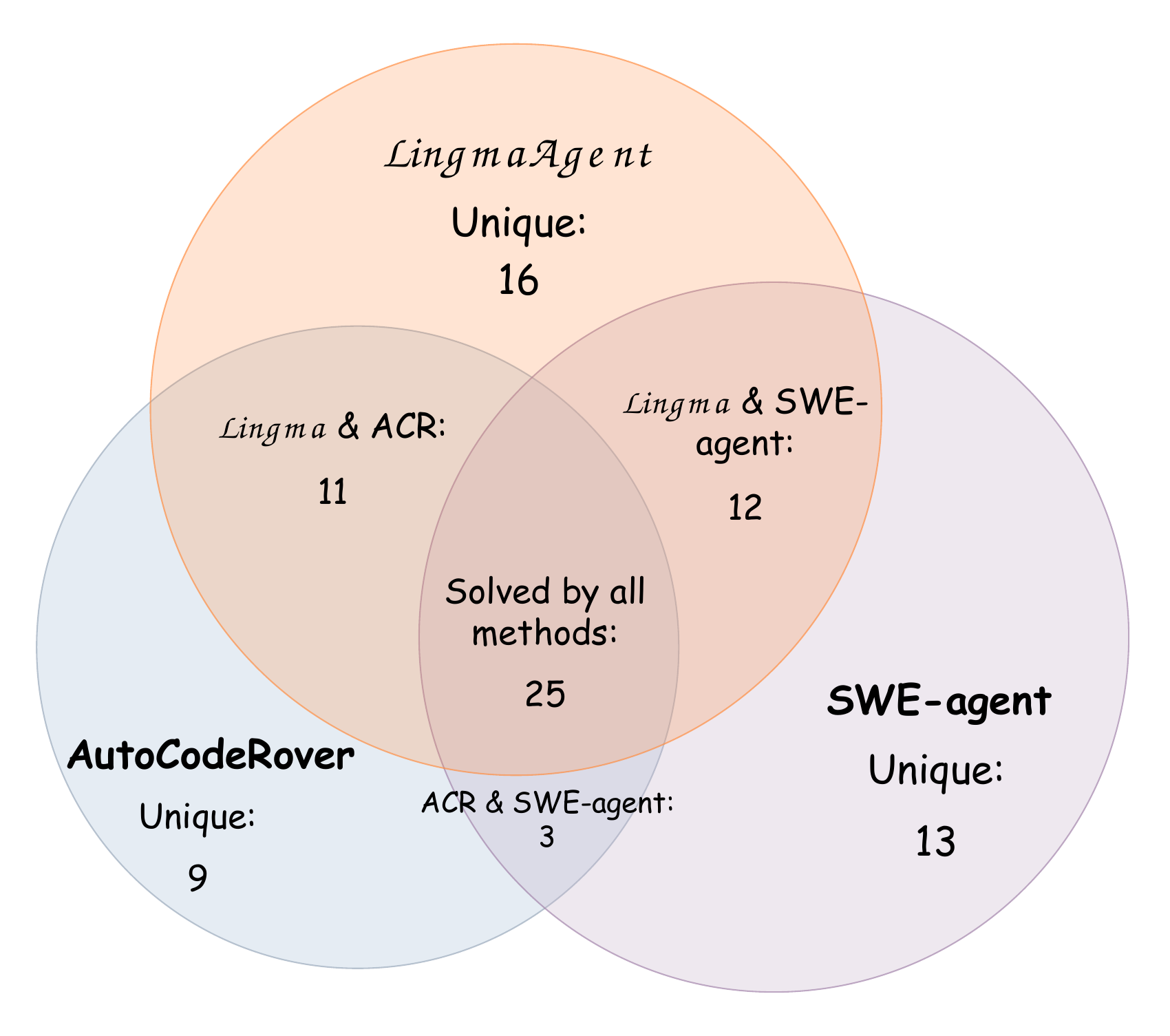}
    \caption{Venn diagrams of resolved cases of Lingma Agent, SWE-agent and AutoCodeRover.}
    \label{fig:result_venn}
\end{figure}

SWE-agent has the highest \textit{Apply} rate due to the introduction of its execution feedback capability. For each issue, SWE-agent first constructs reproduction code to replicate the error, then verifies whether each generated patch has resolved the problem. If not, it uses feedback from the executor to iteratively refine the patch. Because it operates in a real production environment, automatically setting up and obtaining the user's runtime environment may face challenges such as security concerns, environmental complexity, and resource constraints. Therefore, our approach focuses more on understanding the entire repository information. However, to verify the complementarity of the methods, we integrated execution feedback into LingmaAgent.
The detailed results are shown in Table~\ref{tab:venn_diagram_analysis} and Figure~\ref{fig:result_venn}. In Figure~\ref{fig:result_venn}, we compared the issue-solving distribution of three Agent-based methods. We found that our method is highly complementary to the SWE-agent method. The two methods jointly solved 80 examples, achieving a task resolved rate of 26.67\%, which further illustrates the complementarity of our method and the execution feedback method. To further integrate the execution feedback, we followed the practice and prompts of SWE-agent~\citep{yang2024sweagent}. First, we prompted the agent to write code to reproduce the problem, then fix the program and run the reproduction code to determine whether the issue is resolved. If it is not resolved, the agent debugs according to the running results and iteratively refines the generated code to improve the model's output. The experimental result is shown as \textit{LingmaAgent (w.feedback)} in Table~\ref{tab:venn_diagram_analysis}, which we found to be consistent with our expectations and achieved further optimal performance (27.7\%), further verifying the complementarity of LingmaAgent and SWE-agent.

Furthermore, we experimented with different models and found that performance improved significantly with more advanced models. As model capabilities continue to improve and costs decrease, we anticipate even better results in the future. Notably, Claude3.5 Sonnet v1022~\citep{claudev1022} achieved the highest resolution rate of 38.33\% while maintaining a lower average cost, demonstrating the potential for more efficient and effective software engineering problem-solving as LLMs evolve. In addition, We observed that some approaches on SWE-bench leverage voting and test-time scaling~\citep{antoniades2024swe, zhang2408diversity} to enhance performance. However, these methods may introduce significant latency in real-world applications. The exploration of efficient strategies to balance performance gains and latency in practical settings is left for future work.

\subsection{Randomness in LingmaAgent}\label{random}

Product performance stability is crucial for user experience. However, given the inherent randomness in LLMs, it is essential to rigorously assess the consistency of our method's outputs. Therefore, following the practices of AutoCodeRover~\citep{ACR} and SWE-agent~\citep{yang2024sweagent}, we run the system three times to evaluate its average performance and Pass@k performance. These results are shown in Table \ref{tab:random}, where Run$_{1-3}$ represents three different runs, Average represents the average performance of three times, and All represents the result of Pass@3. 

Notably, LingmaAgent consistently outperforms both SWE-agent and  AutoCodeRover across all three runs, with an average improvement of 18.11\% over SWE-agent, the stronger baseline. The Pass@3 performance (represented by "All" in the table \ref{tab:random}) shows that LingmaAgent achieves a success rate of 30.67\%, which is a 12.14\% improvement over SWE-agent and a 37.35\% improvement over AutoCodeRover, indicating its superior ability to solve problems when given multiple attempts. This suggests that the model has the potential to address a wider range of issues, and improving its pass@1 performance could be a promising approach to further enhance its issue-solving capabilities, such as sampling multiple trajectories for DPO/PPO~\citep{rafailov2024direct, wang2024secrets, zheng2023secrets} training, which we leave for future work.

\begin{table}
\renewcommand{\arraystretch}{1.1}
\centering
\scalebox{0.9}{
\begin{tabular}{lccc}
\hline
\textbf{Run} & \textbf{AutoCodeRover} & \textbf{SWE-agent} & \textbf{LingmaAgent}\\
\hline
Run$_1$ & 16.00\%  & 17.33\% & \textbf{21.33\% (23.08\%{\color{red} $\uparrow$})} \\
Run$_2$ & 15.67\% & 18.00\% & \textbf{20.00\% (11.11\%{\color{red} $\uparrow$})} \\
Run$_3$	& 16.67\% & 18.00\% & \textbf{21.67\% (20.39\%{\color{red} $\uparrow$})} \\
Average	& 16.11\% & 17.78\% & \textbf{21.00\% (18.11\%{\color{red} $\uparrow$})} \\
All	& 22.33\% & 27.35\% &  \textbf{30.67\% (12.14\%{\color{red} $\uparrow$})}\\ \hline
\end{tabular}}
\caption{Performance for 3 separate runs of LingmaAgent on SWE bench Lite.}
\label{tab:random}
\end{table}

\subsection{Ablation Study}\label{ablation}

\subsubsection{Module Analysis}\label{module_reason}

This ablation experiment aims to study the effectiveness of LingmaAgent's global repository understanding component. (1) Remove only the call graph module: Only the structure tree in MCTS is retained for tree search, and the reference extension module (i.e., the call relationship graph) is removed. This experiment aims to verify the effectiveness of the reference extension module, i.e., the importance of reference relations in the repository. (2) Remove only the summary module: Only the signature and dependency structure of relevant information in the repository obtained by MCTS are used as global experience, and the summary and planning of information are removed. This experiment aims to verify the effectiveness of the summary agent, i.e., the importance of comprehensive summary of repository information. (3) Remove MCTS \& summary modules: LingmaAgent has no prior knowledge of the repository structure and functions, that is, it lacks empirical information about the whole repository and can only locate relevant code snippets by searching through limited information in the issue. (4) Add a review agent module: After LingmaAgent generates a patch that can be applied, in order to simulate the code review process in the development process, a static review of the patch by the review agent is added to discover possible defects in the newly generated code. If there is a defect, the patch is regenerated according to the review reason until a patch that passes the review is generated. This process is repeated up to three times.

Our experimental results demonstrate the importance of global experience and the effectiveness of the summary agent. As shown in Table \ref{tab:ablation_results}, removing these modules all resulted in a drop in the performance of LingmaAgent, especially after removing the MCTS \& summary agent; the number of problem instances solved rate decreased from 21.33\% to 16.00\%, which highlights the importance of global experience for automatically solving repository-level issues. In addition, we found that after adding the review agent, the performance of LingmaAgent dropped, suggesting the limitations of static review. We speculate that the LLM-based static review may only rely on the surface grammatical information of the code and cannot fully understand the semantic meaning of the code. Therefore, the static review may ignore some hidden logical errors or illogical situations in the code. Therefore, we suggest that subsequent work can combine dynamic program analysis~\citep{zhang2023understanding, deng2023large, xia2024fuzz4all} such as program instrumentation~\citep{hollingsworth1994dynamic, huang1978program} to improve the reliability of the LLM Agent.

\begin{table}
\renewcommand{\arraystretch}{1.3}
\centering
\scalebox{1.0}{
\begin{tabular}{lcc}
\hline
\textbf{Method} & \textbf{Resolved} & \textbf{Apply}\\
\hline
\textbf{LingmaAgent} & \textbf{21.33\%} & 85.67\% \\ 
{ - w.o. call\_graph} & 19.67\% & 83.00\% \\
{  - w/o. summary} & 17.67\% & 85.33\% \\ 
{ - w/o. mcts \& summary} & 16.00\% & 83.33\% \\ 
{  - w. review} & 18.33\% & 87.67\% \\ \hline
\end{tabular}}
\caption{Ablation results of LingmaAgent.}
\label{tab:ablation_results}
\end{table}

\subsubsection{Fault Localization Analysis}\label{fl_reason}

\begin{table}
\renewcommand{\arraystretch}{1.3}
\centering
\scalebox{1.0}{
\begin{tabular}{lcc}
\hline
\textbf{Method} & \textbf{Function\_Loc} & \textbf{File\_Loc}\\
\hline
AutoCodeRover & 42.3\% & 62.3\% \\  
SWE-agent & 45.3\% & 61.0\% \\  
\textbf{LingmaAgent} & \textbf{49.3\%} & \textbf{67.7\%} \\  \hline
\end{tabular}}
\caption{Fault localization results of LingmaAgent.}
\label{tab:fl_results}
\end{table}

In addition to the issue resolution evaluation, we conducted a analysis focusing on the fault localization capability (\% Correct Location)~\citep{agentless} of LingmaAgent. The fault localization module is an intermediate module in the issue solving process. Whether the fault location is correctly located affects the subsequent program repair. Specifically, we extracted the fault locations from the developer patch and the patch generated by the model, respectively, and calculated the localization success rate by calculating whether the fault locations were consistent. This analysis aims to further illustrate the effectiveness of our repository understanding module. The resluts are shown in Table \ref{tab:fl_results}. We compared the two SOTA agent-based methods, AutoCodeRover and SWE-agent, where Function represents the accuracy of fault function location and File represents the accuracy of defect file location. Our findings show that our method significantly outperforms the other two methods in the success rate of fault localization at both the Function and File levels, which shows that understanding the repository and exploring critical information notably contribute to improving fault localization.

\subsubsection{Hyper Parameter Analysis}\label{parameter_reason}

\begin{table}
\renewcommand{\arraystretch}{1.2}
\centering
\scalebox{1.05}{
\begin{tabular}{ccc}
\hline
\textbf{MCTS\_Iters} & \textbf{Resolved} & \textbf{Apply}\\
\hline
0 & {16.00\% (48)} & 80.33\% \\
50 & {19.67\% (59)} & 86.67\% \\ 
200 & {20.67\% (62)} & 88.00\% \\ 
\textbf{600} & \textbf{21.33\% (64)} & 85.67\% \\ 
\hline
\end{tabular}}
\caption{Hyperparameter results.}
\label{tab:hyperparameter_results}
\end{table}

We further analyzed the impact of the iterations number in MCTS. We set the maximum number of iterations to 50, 200, and 600, and limited the maximum iteration time to 300 seconds. The results are shown in Table \ref{tab:hyperparameter_results}. We found that: (1) As the number of iterations increases, LingmaAgent solves more actual issues. This shows that as the number of iterations rounds increases, agents will collect more repository information, i.e., they will have more experience with the repository, resulting in a higher problem solving rate; (2) As the number of iterations increases, we found that the relative improvement in problem solving gradually decreases. Specifically, the improvement of 50 iterations is significant compared to no iterations, but the relative improvement of the subsequent 200 and 600 iterations decreases. This may be because in the early stage, agents can quickly search and summarize relevant experience, but as the number of iterations increases, the convergence speed of the model gradually slows down, and the contribution of new information to performance improvement becomes smaller; (3) We observed that as the number of iterations increases from 200 to 600, the apply rate decreases. This phenomenon indicates that as the number of iterations increases, the model may be affected by some interference information when generating results, resulting in a decrease in the quality of the generated results. Therefore, when selecting the number of iterations, it is necessary to consider avoiding the influence of excessive interference information.

\subsection{Evaluation on In-house Dataset}\label{inhouse}

\begin{table}
\renewcommand{\arraystretch}{1.2}
\centering
\scalebox{1.05}{
\begin{tabular}{lcc}
\hline
\textbf{Language} & \textbf{Resolved} & \textbf{Fault\_Location}\\
\hline
Java & 14.7\% & 41.2\% \\ 
TypeScript & 18.8\% & 28.1\% \\ 
JavaScript & 17.2\% & 31.3\% \\ 
\textbf{Average} & \textbf{16.9\%} & \textbf{33.5\%} \\
\hline
\end{tabular}}
\caption{Results on Multilingual In-house Dataset.}
\label{tab:in_house_eval_mul}
\end{table}

\begin{table}
\renewcommand{\arraystretch}{1.2}
\centering
\scalebox{1.0}{
\begin{tabular}{lcc}
\hline
\textbf{Tasks} & \textbf{Automated} & \textbf{Human-in-the-Loop}\\
\hline
Resolved & 16.7\% & \textbf{43.3\%} \\
Fault\_Loc & 40.0\% & \textbf{66.7\%} \\ \hline
\end{tabular}}
\caption{Results of Human-in-the-Loop Intervention on Alibaba Cloud In-house Dataset Subset.}
\label{tab:in_house_eval}
\end{table}

To assess the effectiveness of LingmaAgent in real-world industrial settings, we conducted a comprehensive evaluation using an in-house dataset meticulously curated from Alibaba Cloud's diverse development scenarios. This dataset was designed to test the performance of LingmaAgent on multi-language repositories, focusing on Java, JavaScript, and TypeScript - three of the most prevalent languages in cloud-based and web application contexts. The dataset encompasses a wide range of projects including e-commerce platforms, cloud infrastructure services, and data analytics tools, representing the complexity and diversity of industrial-scale software projects. It comprises 10 Java repositories (averaging 1538 files and 3.4 issues each), 24 JavaScript repositories (averaging 503 files and 4 issues each), and 16 TypeScript repositories (averaging 793 files and 4 issues each), for a total of 194 issues.

We deployed LingmaAgent with the GPT-4 for this evaluation. We used the same metrics to evaluate LingmaAgent's performance. The experiment was conducted in two phases:
\begin{itemize}
\item Fully Automated Resolution. LingmaAgent attempted to resolve issues without any human intervention. For this phase, we conducted a full evaluation on all 194 issues in the dataset.
\item Human-in-the-Loop Intervention. For issues not resolved in the first phase, we implemented a human-in-the-loop approach. Development engineers intervened in the product interaction pipeline, manually adjusting LingmaAgent's generation plans and search\_api call, and refining potential fault localizations (interventions must less than 5 times). The final patches were then generated using the model. For this phase, we randomly selected 30 issues for manual evaluation.
\end{itemize}

The results of our evaluation are presented in Table \ref{tab:in_house_eval_mul} and \ref{tab:in_house_eval}. These findings offer valuable insights into LingmaAgent's performance in industrial-scale software engineering tasks. In the Fully Automated Resolution phase, LingmaAgent successfully resolved 16.9\% of the issues across the multilingual dataset. This result shows that the system is capable of autonomously handling some real software engineering tasks without any human intervention, but there is still much room for improvement. For the Human-in-the-Loop Intervention phase, we randomly selected a subset of 30 issues. In this subset, LingmaAgent's automated performance was consistent with the full dataset, resolving 16.7\% of issues independently. However, with human intervention, the resolution rate dramatically increased to 43.3\%. This substantial improvement of 26.7 percentage points highlights the synergistic potential of human-AI collaboration in tackling complex software engineering problems. This indicates that the system effectively augments human problem-solving skills rather than replacing them.
Overall, these results demonstrate LingmaAgent's potential as a powerful tool in industrial software engineering contexts, particularly when integrated into a collaborative workflow with human developers.

\section{Limitation}





\textbf{Resource Overhead.} Although LingmaAgent aims to guide LLMs to fully understand the whole software repository to effectively solve the challenges in ASE, the MCTS process does require a certain amount of resource consumption. Specifically, we set the maximum number of iterations to 600 and the maximum search time to 300 seconds to ensure that the model can fully explore the search space and accurately evaluate the rewards of different paths. However, such settings are controllable and adjustable to adapt to different application scenarios and resource constraints. Through reasonable parameter adjustment, the best balance between resource consumption and result accuracy can be found. In addition, as shown in Table \ref{tab:hyperparameter_results}, only 50 iterations can also achieve results that are superior to other agents. At the same time, in Table \ref{tab:venn_diagram_analysis}, we ran LingmaAgent on different base models. We found that with the introduction of the next generation of models, the improvement of model capability and the cost will also bring about the improvement of LingmaAgent effect and the reduction of cost, which further demonstrates the possibility of application with LingmaAgent. Further research may discover more efficient strategies to reduce resource requirements while maintaining or improving agents performance.

\textbf{Evaluation of LingmaAgent.} While our evaluation of LingmaAgent demonstrates promising results, several limitations in our current assessment approach warrant consideration. Our primary evaluation relies on the SWE-bench Lite dataset, and although we conducted additional tests using an in-house dataset from Alibaba Cloud, the scope of our human-in-the-loop evaluation was limited due to the substantial human resources required for comprehensive interaction and assessment. This constraint potentially limits the generalizability of our findings to a broader range of real-world scenarios. Additionally, there is a possibility that some of the models we used, may have been exposed to parts of the repositories in our test set during their training. While this potential data leakage is a concern, we believe that the relative performance improvements demonstrated by LingmaAgent still provide valuable insights into its effectiveness. Nevertheless, this limitation highlights the need for more controlled evaluation environments in future studies. To address these limitations and enhance the robustness of future evaluations, we propose several directions for future work: creating standardized protocols for human-in-the-loop evaluations, and developing a dynamic, continuously updated version of SWE-bench that evolves with the software engineering field. By addressing these limitations and expanding our evaluation methodologies, we aim to provide more robust and generalizable assessments of AI-assisted software engineering tools like LingmaAgent in the future.



\section{Conclusion}
This paper emphasizes the importance of understanding entire software repositories for achieving Automated Software Engineering. We introduce LingmaAgent, a novel LLM-based agent method that comprehensively analyzes repositories through knowledge graph construction, MCTS-enhanced exploration, and global experience-based planning. This approach enables agents to solve real-world GitHub issues effectively. Extensive experiments demonstrate LingmaAgent's superior performance over existing systems on the SWE-bench Lite benchmark. Ablation studies highlight the significance of global repository experiences and the potential of integrating runtime feedback. We also validate LingmaAgent's effectiveness in real-world industrial settings using an Alibaba Cloud dataset, showcasing its capabilities in both automated and human-in-the-loop scenarios.

Future work will focus on developing a dynamic, evolving version of SWE-bench, and optimizing resource efficiency while maintaining or improving agent performance. These efforts aim to advance the field of AI-assisted software engineering and provide more robust solutions for complex software development tasks.

\bibliographystyle{ACM-Reference-Format}
\bibliography{sample-base}


\end{document}